\documentclass[twocolumn,aps,prd,superscriptaddress,nofootinbib,preprintnumbers]{revtex4-1}
\usepackage{graphicx}
\usepackage{amsmath}
\usepackage{bm}
\usepackage{slashed}
\usepackage{epsfig}
\usepackage{amsfonts}
\usepackage{epstopdf}
\usepackage{hyperref}
\usepackage[subfigure]{graphfig}
\usepackage{braket}
\usepackage{tikz}
\usepackage{hhline}
\usepackage[normalem]{ulem} 
\usepackage{soul}

\newcommand{\ben}{\begin{eqnarray}}
\newcommand{\een}{\end{eqnarray}}

\newcommand{\bef}{\begin{figure}[!htp]}
\newcommand{\eef}{\end{figure}}

\newcommand{\nn}{\nonumber}

\newcommand{\om}{{\omega}}

\def\be{\begin{equation}}
\def\ee{\end{equation}}
\newcommand{\bea}{\begin{eqnarray}}
\newcommand{\eea}{\end{eqnarray}}

\def\ba{\begin{linenomath*}\begin{equation}}
\def\ea{\end{equation}\end{linenomath*}}

\usepackage{soul}

\hypersetup{colorlinks, linkcolor = [rgb]{0,0.0,0.5}, citecolor = [rgb]{0,0.0,0.5}, urlcolor = [rgb]{0,0.0,0.5}}

\begin{document}


\title{ Gluon  distributions and  their applications to Ioffe-time distributions}

\newcommand*{\WM}{Physics Department, William \& Mary, Williamsburg, Virginia 23187, USA}\affiliation{\WM}
\newcommand*{\JLAB}{Thomas Jefferson National Accelerator Facility, Newport News, VA 23606, USA}\affiliation{\JLAB}
\newcommand*{\SDU}{Key Laboratory of Particle Physics and Particle Irradiation (MOE), Institute of Frontier and Interdisciplinary Science, Shandong University, Qingdao, Shandong 266237, China}\affiliation{\SDU}
\newcommand*{\ICTP}{The Abdus Salam International Centre for Theoretical Physics, Strada Costiera 11, I-34014,Trieste, Italy}\affiliation{\ICTP}

\author{Raza Sabbir Sufian}\email{Corresponding author: sufian@jlab.org}\affiliation{\WM}\affiliation{\JLAB}
\author{Tianbo Liu}\email{Corresponding author: liutb@sdu.edu.cn}\affiliation{\SDU}\affiliation{\JLAB}
\author{Arpon Paul}\affiliation{\ICTP}

\begin{abstract}
We investigate unpolarized and polarized gluon distributions and their applications to the Ioffe-time distributions, which are related to lattice QCD calculations of parton distribution functions. Guided by the counting rules based on the perturbative QCD at large momentum fraction $x$ and the color coherence of gluon couplings at small $x$, we parametrize gluon distributions in the helicity basis. By fitting the unpolarized gluon distribution, the inferred polarized gluon distribution from our parametrization agrees with the one from global analysis. A simultaneous fit to both unpolarized and polarized gluon distributions is also performed to explore the model uncertainty. The agreement with the global analysis supports the $(1-x)$ power suppression of the helicity-antialigned distribution relative to the helicity-aligned distribution. The corresponding Ioffe-time distributions and their asymptotic expansions are calculated from the gluon distributions. Our results of the Ioffe-time distributions can provide  guidance to the extrapolation of lattice QCD data to the region lacking precise gluonic matrix elements. Therefore, they can help regulate the ill-posed inverse problem associated with extracting the gluon distributions from discrete data from first-principle calculations, which are available in a limited range of the nucleon momentum and the spatial separation between the gluonic currents. Given various limitations in obtaining lattice QCD data at 
large Ioffe time, phenomenological approaches can provide important complementary information  to extract the gluon distributions in the entire 
momentum fraction region, especially at small $x$. The possibility of investigating higher-twist effects and other systematic uncertainties in the contemporary first-principle calculations of parton distributions from phenomenologically well-determined Ioffe-time distributions in the large Ioffe-time region is also discussed.

\end{abstract}

\maketitle
\allowdisplaybreaks


\section{Introduction}
One of the outstanding  problems in  nuclear and particle physics is  to understand the structure of hadrons in terms of quarks and gluons, the fundamental degrees of freedom in QCD.  Gluons, which serve as  mediator bosons of the strong interaction while also carrying the color charge, play a key role in the nucleon's mass and spin.  With great progress in the extraction of nucleon parton distribution functions (PDFs) in the past decades, especially the quark distributions, the understanding of the gluon distribution and its role in hadron structures remains one of the most challenging but fundamental issues in nuclear and particle physics.

 Given the fact that no free quarks or gluons have been observed due to the confinement, most analyses of hadron involved high energy scatterings rely on the QCD factorization, where PDFs play an important role. Compared to quark distributions, the gluon distribution is less accurately extracted, which may affect the calculation of the cross section of a process dominated by the gluon-fusion channel, {\it e.g.} the Higgs boson production at the LHC~\cite{Georgi:1977gs}. While the precision of the extracted $g(x)$ has been improved a lot during the last decade, there are still some issues like the suppression in the momentum fraction region $ 0.1 < x < 0.4$ when ATLAS and CMS jet data are included~\cite{Hou:2019efy}. Obtaining a more precise determination of $g(x)$ is subject to ongoing efforts in global analyses of PDFs.  In contrast to the unpolarized PDFs, the polarized PDFs, especially the polarized gluon distribution $\Delta g(x)$ as well as sea quark distributions, are poorly determined, even the sign is not fully determined.  One of the main physics goals of the upcoming Electron-Ion-Collider (EIC)~\cite{Accardi:2012qut} is to have precise measurements of the nucleon spin structure, particularly the gluon and sea quark distributions.  

It has been 30 years since the EMC experiment~\cite{Ashman:1987hv}, which found that only a small fraction of the nucleon spin is carried by the quark spin and triggered the so-called proton spin puzzle. The remaining part of the proton spin is usually assigned to the orbital angular momenta and the gluon spin. After significant efforts in the last decades, the quark spin part was found to contribute only about 30\% to the proton spin. Recent experimental data from RHIC and lattice QCD calculation suggest that gluon may contribute a large amount to the proton spin. For a recent review, see~\cite{Ji:2020ena}. There have been several global analyses using different experimental data sets and different types of parametrizations~\cite{Gehrmann:1995ag,Gluck:2000dy,Bluemlein:2002be,Leader:2005ci} to impose constraints on $\Delta G$. A recent extraction with updated data sets and PHENIX measurement~\cite{Adare:2008aa} of double helicity asymmetry in inclusive $\pi^0$ production in polarized $p-p$ collision obtained $\Delta G=0.2$ with a constraint of $-0.7 < \Delta G < 0.5$ for the sampled gluon momentum fraction $x\in [0.02, 0.3]$. Excluding the $x<0.05$ region, the value of $\Delta G =\int_{0.05}^{0.2} dx~\Delta g(x)=0.23(6)$~\cite{Nocera:2014gqa} and $\Delta G =\int_{0.05}^{1} dx~\Delta g(x)=0.19(6)$~\cite{deFlorian:2014yva} were obtained. Future experimental measurement of $\Delta g(x)$ in the $x<0.02$ is required to reduce the uncertainty in $\Delta G$. We note that $\Delta g(x)$ extracted mostly from the double longitudinal spin asymmetry is always limited to some $x_{\rm min}$ no matter how high the energy of the experimental set up is and some theoretical calculations are required to constrain $\Delta g(x)$ at low $x$~\cite{Kovchegov:2017lsr}. Fortunately, one of the major goals of the upcoming EIC~\cite{Accardi:2012qut} is to precisely explore $\Delta g(x)$ at  low $x$ and provide  stringent constraints on the gluon helicity distribution.

Since $\Delta G$ is not related to the local matrix element of a gauge invariant operator, it could not be directly calculated in the lattice QCD (LQCD) calculations. However, following the formalism proposed in~\cite{Ji:2013fga}, it has become possible to  calculate $\Delta G$ in terms of a local and gauge invariant operator in LQCD. Since then,  there has only been one direct LQCD calculation~\cite{Sufian:2014jma,Yang:2016plb} of the gluon spin content in the nucleon. With leading-order matching using Ji's large-momentum effective theory (LaMET)~\cite{Ji:2013dva,Ji:2014gla}, it was determined that \mbox{$\Delta G (\mu^2=10{\rm GeV}^2)=0.251(47)(16)$}, {\it i.e.} about 50\% of the proton spin comes from the gluons. However,  a refined study  with the investigation of the convergence of matching beyond 1-loop, the estimate of power corrections, and other sources of systematic uncertainties are warranted to obtain an unambiguous determination of $\Delta G$ in the future LQCD  calculations. 

Now, the PDFs of the quarks and gluons contain the nonperturbative structure of hadrons, especially at the low-resolution  scale $Q^2$. However, the possibility that even at low $Q^2$ one can obtain a reasonable shape and distribution of the hadron structure functions by transcribing our knowledge of the perturbative QCD (pQCD) based counting rules~\cite{Brodsky:1973kr} as $x\to 1$ and color coherence of gluon couplings as $x\to 0$ has been shown to provide promising outcomes in many theoretical calculations.  For example, calculations of the unpolarized and polarized quark and gluon PDFs in~\cite{Brodsky:1989db,Brodsky:1994kg} showed practical application of these limiting behaviors of PDFs by obtaining PDFs in agreement with the analysis in~\cite{Martin:1992zi}. The momentum fraction carried by the gluon in the nucleon $\langle x\rangle_g\approx 0.42$ determined in~\cite{Brodsky:1994kg} is in remarkable agreement with recent global analyses~\cite{Harland-Lang:2014zoa,Dulat:2015mca,Ball:2017nwa}. To emphasize further, most of the earlier and present-day global analyses~\cite{Harland-Lang:2014zoa,Dulat:2015mca,Alekhin:2017kpj,Ethier:2017zbq} use some functional forms similar to $x^\alpha(1-x)^\beta\mathcal{F}(x)$ where the asymptotic behavior of the PDFs at small $x$ ($x^\alpha$ behavior) is adopted from the observed Regge behavior~\cite{Regge:1959mz} in particle colliders and the large-$x$ behavior ($(1-x)^\beta$ fall-off)  based on the power counting rules  for hard scattering ~\cite{Brodsky:1973kr} with some interpolating function $\mathcal{F}(a_i,b_i,\cdots,x)$ with unknown parameters $a_i,b_i,\cdots$ between these two limits that varies in different parametrizations of PDFs. Once these PDFs are determined at some initial scale, their $Q^2$-evolution is well predicted in pQCD through the DGLAP equation~\cite{Gribov:1972ri,Altarelli:1977zs,Dokshitzer:1977sg}. These PDFs determined at the low initial scale have been shown to be universal between different reactions with their scale-dependent modifications governed by pQCD evolution,  and therefore indicated  that these nonperturbative universal PDFs can indeed  be  well approximated even at low $Q^2$ by incorporating pQCD constraints at large $x$ and Regge behavior at small $x$.  Very good agreement and consistent behavior of the nucleon unpolarized  PDFs and precise prediction of nucleon polarized distributions from the unpolarized PDFs  have also been possible in recent calculations~\cite{deTeramond:2018ecg,Liu:2019vsn} where the PDFs  are governed by these limiting behaviors. Similarly, recent synergies between LQCD and phenomenological calculations have provided useful constraints in the study sea-quark asymmetry in the nucleon with higher precision than either theory or experiment alone  could attain~\cite{Sufian:2018cpj,Sufian:2020coz}.

In light of the above discussions, we revisit the calculations in~\cite{Brodsky:1989db,Brodsky:1994kg} which incorporated pQCD constraints at large $x$ and coherent correlations of partons at low $x$ to determine unpolarized and polarized gluon distributions. These calculations~\cite{Brodsky:1989db,Brodsky:1994kg} demonstrated that many properties of the exclusive reactions  can be calculated  by incorporating the knowledge of asymptotic freedom, power-law scaling, and helicity conservation rules of pQCD  without explicit knowledge of the nonperturbative light-front wave function.  

The main goal of this article is to transcribe these insights from the small and large $x$ physics and compare how adequate and compatible they are with the recent determinations of gluon distributions.  We emphasize  this calculation does not aim to provide the precise determination of the gluon PDFs, rather our main focus is to determine the shapes of the gluon PDFs based on these simple constraints in the small and large $x$-regions.  Here we point out that, we do not focus here on the important aspects of gluon distributions at extremely small $x$-values which has been discussed in the literature~\cite{Kuraev:1977fs,Balitsky:1978ic,Kirschner:1983di,Gribov:1984tu,Mueller:1985wy,Balitsky:1995ub,Kovchegov:1999yj}. Another important goal of the upcoming EIC is to explore the very low-$x$ region where saturation of gluon densities sets in~\cite{McLerran:1993ni,JalilianMarian:1996xn} and has not yet been conclusively observed. Parton distributions at extremely small-$x$ is an active field of research and we avoid the discussion of the related complication here. We first determine the unknown coefficients in the parametrization of helicity aligned $g^+(x)$ and anti-aligned $g^-(x)$ gluon distributions using the global fits of unpolarized gluon distribution and use those to  calculate the  polarized gluon distribution and  gluon asymmetry distribution $\Delta g(x)/g(x)$. We calculate the corresponding Ioffe-time distributions (ITDs)~\cite{Gribov:1965hf,Ioffe:1969kf,Braun:1994jq} of the unpolarized and polarized gluon  distributions and demonstrate how these can provide valuable information and important constraints in the determination of full $x$-dependence of PDFs and also  their higher moments in the future LQCD calculations. In particular, we determine the  asymptotic behavior of the unpolarized and polarized gluon ITDs which are not accessible within the  current reach of  LQCD calculations and can provide complementary information to reconstruct the full $x$-dependence of the unpolarized and polarized gluon distributions.


\section{Gluon distributions from helicity-basis parametrization}\label{gluonPDFs}

To construct the parametrization of the helicity-basis gluon distributions, $g^+(x)$ for helicity-aligned distributions and $g^-(x)$ for helicity-antialigned distribution, we consider the counting rules based on perturbative QCD analysis~\cite{Brodsky:1994kg}. Compared to valence quark distributions, which fall off as $(1-x)^3$ as $x\to1$, $g^+(x)$ is suggested to fall-off faster as $(1-x)^4$ and $g^-(x)$ is expected to be further suppressed by $(1-x)^2$.
 Although the 
$(1-x)$ power behavior qualitatively provides the fall-off feature of 
the distributions at large $x$, the exact power values depend on the 
scale, which is not specified in the perturbative QCD analysis~\cite{Brodsky:1994kg}.
Instead of strictly imposing the power counting  as $x\to1$ and the Pomeron intercept  as $x\to0$, we only take them as  guidance and phenomenologically introduce two parameters $\alpha$ and $\beta$ to allow the variation of the power behavior at small and large $x$ regions as usually adopted in global analyses. For a  good description of the gluon distribution in the full-$x$ region, we also include a polynomial $(1+\gamma\sqrt{x}+\delta x)$ with parameters $\gamma$ and $\delta$ to be fitted. As a modification of the  functional form utilized in~\cite{Brodsky:1994kg} by including the polynomial, we parametrize the  helicity-aligned and the  helicity-antialigned gluon distributions as
\bea \label{a1}
xg^+(x)&=& x^\alpha\big[A(1-x)^{4+\beta}+B(1-x)^{5+\beta}\big]\nn \\
&&\times(1+\gamma\sqrt{x}+\delta x),\nn \\
xg^-(x)&=& x^\alpha\big[A(1-x)^{6+\beta}+B(1-x)^{7+\beta}\big]\nn \\
&&\times(1+\gamma\sqrt{x}+\delta x), 
\eea
where $A$ and $B$ are normalization parameters to be determined. The inclusion of the subleading term in the power of $(1-x)$ is to account for the contribution from higher Fock state. For each term, the power of $(1-x)$ differs by $2$ as suggested by the pQCD analysis~\cite{Brodsky:1994kg}. We refer to the parametrization form Eq.~\eqref{a1} as the ansatz-1.

As a phenomenological exploration,  we consider another option of $g^-(x)$ being  suppressed by one power of $(1-x)$ in comparison with the $g^+(x)$. This results in the parametrization,
\bea \label{a2}
xg^+(x)&=& x^\alpha\big[A(1-x)^{4+\beta}+B(1-x)^{5+\beta}\big]\nn \\
&&(1+\gamma\sqrt{x}+\delta x),\nn \\
xg^-(x)&=& x^\alpha\big[A(1-x)^{5+\beta}+B(1-x)^{6+\beta}\big]\nn \\
&&(1+\gamma\sqrt{x}+\delta x), 
\eea
which we refer to as the ansatz-2.  As we will discuss later, fixing the $(1-x)$ power difference between $g^+(x)$ and $g^-(x)$ introduces a model bias, which leads to an underestimation of the uncertainties. To investigate the model uncertainty, we consider a more flexible parametrization,
\bea \label{asim}
xg^+(x)&=& x^\alpha\big[A(1-x)^{4+\beta}+B(1-x)^{5+\beta}\big]\nn \\
&&(1+\gamma\sqrt{x}+\delta x),\nn \\
xg^-(x)&=& x^\alpha\big[A(1-x)^{6+\beta'}+B(1-x)^{7+\beta'}\big]\nn \\
&&(1+\gamma'\sqrt{x}+\delta' x), 
\eea
 where the $(1-x)$ exponents and the polynomial coefficients in $g^+(x)$ and $g^-(x)$ are independent parameters. We refer to this parametrization as ansatz-3.
 We note that all these ansatzes have the $\Delta g(x)$ approaching to $0$ as $x\to0$. This indicates that the helicity correlation between the gluon and its parent nucleon disappears when $x\to0$, where the relative rapidity becomes infinity. The saturation effect may suppress the evolution of helicity distributions at small-$x$ and consequently leave a small amount of spin contribution in the small-$x$ region~\cite{Kovchegov:2017lsr,Kovchegov:2016zex}. Since the goal of this paper is not the small-$x$ distribution, we limit to the assumption above in this study and restrict the subsequent  analyses in the $x\geq 10^{-3}$ region.

With the parametrization of the  helicity-aligned and the  helicity-antialigned gluon distributions, one can directly obtain the unpolarized and polarized gluon distributions from the sum and the difference of them,
\bea \label{a3}
xg(x) &\equiv& xg^+(x)+xg^-(x),\\ \label{3}
x\Delta g(x) &\equiv & xg^+(x)-xg^-(x).\label{a4}
\eea

\begin{figure}[htp]
\begin{center}
\includegraphics[width=3.45in, height=2.5in]{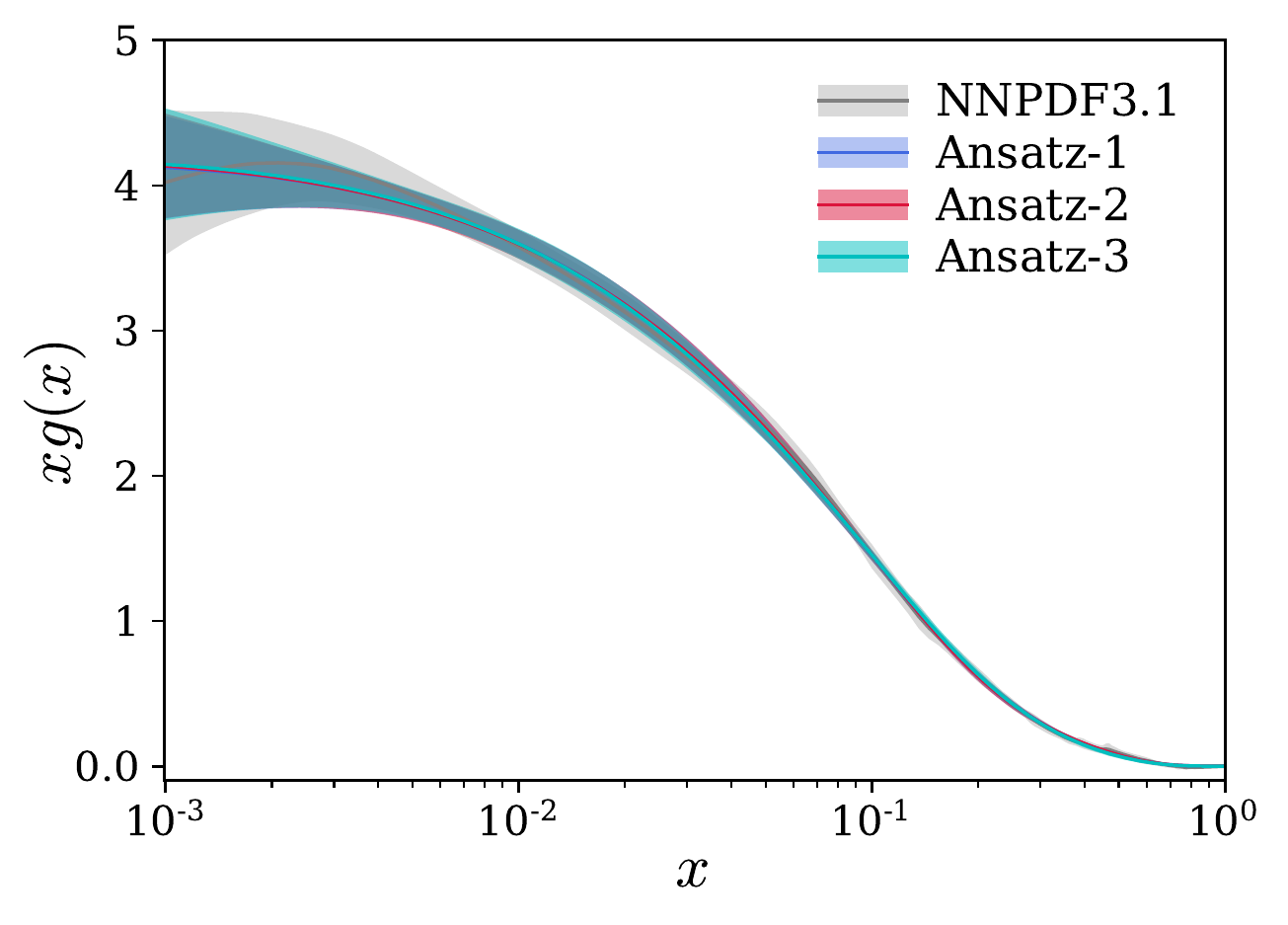}
\caption{\label{fig:1}
 Unpolarized gluon distributions obtained by fitting ansatz-1  (Eq.~\eqref{a1}), ansatz-2 (Eq.~\eqref{a2}), and ansatz-3 (Eq.~\eqref{asim}) to the NNPDF distribution at the factorization scale $\mu=2$ GeV. The gray band shows the unpolarized gluon distribution $xg(x)$ as given by the NNPDF global analysis. The blue, red, and  cyan bands labeled by Ansatz-1, Ansatz-2, and Ansatz-3 show distributions determined according to  ans{\"a}tze-1, 2, and 3 for the $xg^+(x)$ and $xg^-(x)$ distributions, respectively.} 
\end{center}
\end{figure}

To determine the  parameters  in ansatz-1 and ansatz-2, we fit the unpolarized gluon distribution from the NNPDF global analysis~\cite{Ball:2017nwa} at the factorization scale $\mu=2$ GeV. Our procedure described here can be applied to any other gluon distribution given by global analyses~\cite{Harland-Lang:2014zoa,Dulat:2015mca,Alekhin:2017kpj,Ethier:2017zbq} or model calculation. To fit the distribution in the full-$x$ range, we select 200 points in $x$ values. 100 of them are equally separated from $10^{-4}$ to  $10^{-1}$ in the logarithmic scale and the other 100  $x$-values are equally separated from $10^{-1}$ to 1 in the linear scale. Each point is weighted by the inverse square of its uncertainty given by the global analysis. We take the 100 replicas of the gluon distribution from NNPDF3.1 NLO PDF  set~\cite{Ball:2017nwa}. For each replica, we perform a fit to determine the parameters. In the end, we have 100 sets of parameters, which determine the gluon distributions following the ansatz-1 or the ansatz-2.  For the ansatz-3 in which the parameters $\beta$, $\gamma$ and $\delta$ are chosen independently for $g^+(x)$ and $g^-(x)$, we perform a simultaneous fit to the  unpolarized~\cite{Ball:2017nwa} and polarized~\cite{Nocera:2014gqa} gluon distributions. In this case, the result of the polarized gluon distribution is driven by the global fit. As a result, the polarized gluon distribution associated with ansatz-3 has a better match with the NNPDF global analysis. The results of the unpolarized gluon distribution are compared with the global analysis in FIG.~\ref{fig:1}, where the central value is evaluated from the average value of the 100 replicas for each ansatz and the uncertainty band is the standard deviation among them. One can observe that the three ansatzes have almost indifferentiable results and match the global analysis well.  For completeness, we list the fitted values of the parameters in Table~\ref{tab:power}.

\begin{table}[htp]
    \centering
    \caption{The fitted values of parameters in the three parametrization ansatzes. The second row of ansatz-3 gives the values of parameters $\beta'$, $\gamma'$, and $\delta'$.}
    \label{tab:power}
    \begin{tabular}{lcccc}
    \hline\hline
    Ansatz   & ~~~~~$\alpha$~~~~~ & ~~~~~$\beta$~~~~~ & ~~~~~$\gamma$~~~~~ & ~~~~~$\delta$~~~~~ \\
    \hline
    1 & $0.036\pm0.058$ & $0.95\pm1.28$ & $-2.80\pm0.63$  & $2.62\pm0.95$  \\
    2 & $0.034\pm0.060$ & $1.11\pm1.32$ & $-2.87\pm0.56$  & $2.67\pm0.86$  \\
    3 & $0.034\pm0.064$ & $0.54\pm1.30$ & $-2.63\pm0.60$  & $2.54\pm1.01$  \\
      &       ---       & $0.91\pm2.63$ & $-2.55\pm0.95$  & $3.24\pm2.83$  \\
    \hline\hline
    \end{tabular}
\end{table}

\begin{figure}[htp]
    \centering
			\includegraphics[width=3.45in, height=2.5in]{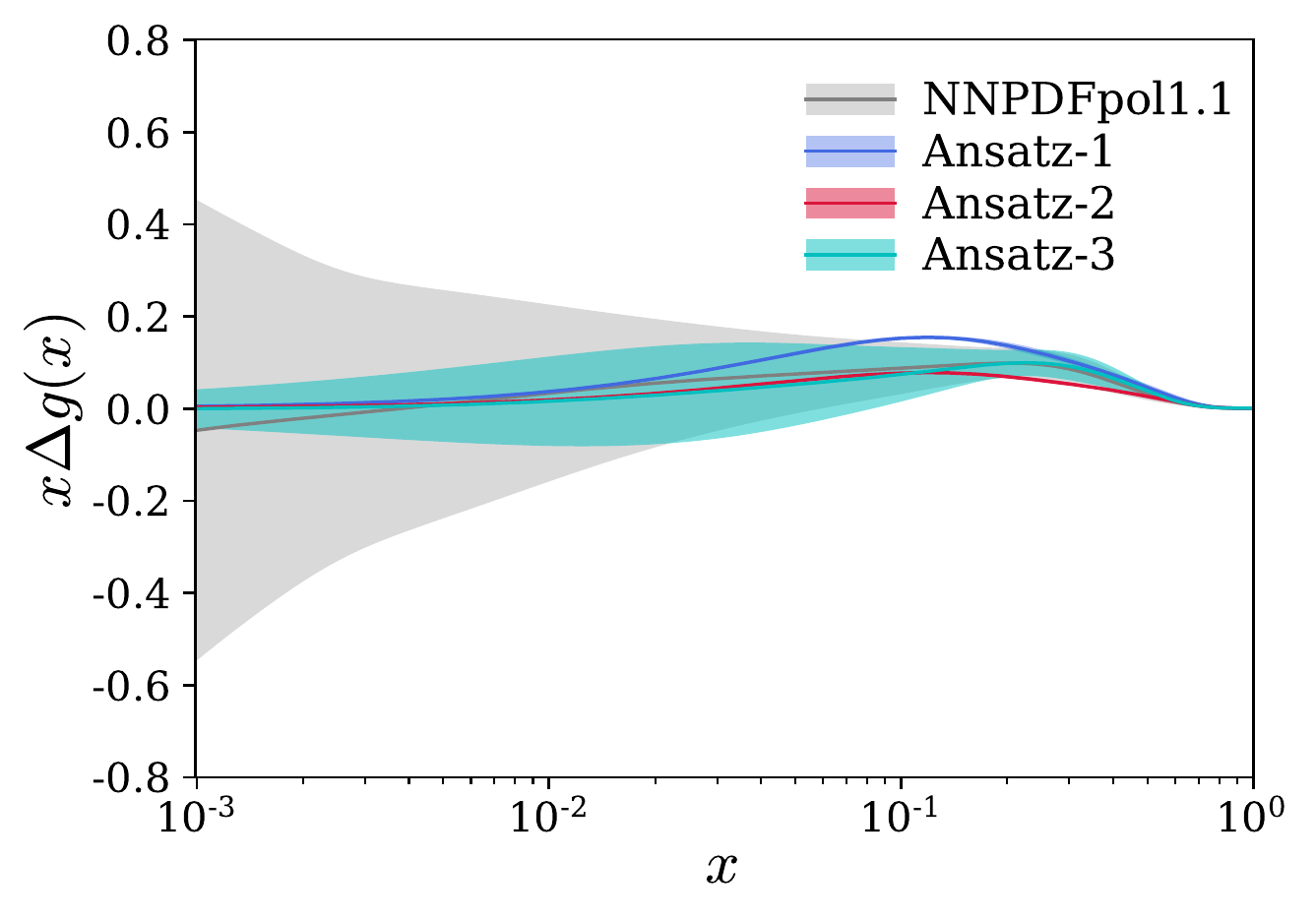}
			\caption{\label{fig:2}  Polarized gluon distributions from the fit parameters determined from fitting ansatz-1 and ansatz-2 to the NNPDF unpolarized gluon distribution.  Ansatz-3 refers to the polarized gluon distribution obtained from a simultaneous fit to the NNPDF3.1 NLO PDF  set~\cite{Ball:2017nwa} and NNPDFpol1.1 PDF set~\cite{Nocera:2014gqa} using fit paramterization in Eq.~\eqref{asim}. The gray band shows the polarized gluon distribution $x\Delta g(x)$ as given by the NNPDFpol1.1 global analysis~\cite{Nocera:2014gqa}. The  blue, red, and cyan bands labeled by Ansatz-1, Ansatz-2, and Ansatz-3 show distributions  obtained using  parameters obtained in the fits of $xg^+(x)$ and $xg^-(x)$ to the  NNPDF gluon distribution.}
\end{figure}

From the definition of the polarized gluon distribution in Eq.~\eqref{a4}, we now obtain the polarized gluon distribution based on the above fit results of $xg^+(x)$ and $xg^-(x)$. Unlike the unpolarized distribution, the results of $\Delta g(x)$ determined from  ansatz-1, 2, and 3 have observable difference, especially in the region $10^{-2}\lesssim x\lesssim 0.5$, as shown in FIG.~\ref{fig:2}. However, all these  determinations of $x\Delta g(x)$ are in good agreement within the uncertainty range of the NNPDF global analysis  with a noticeable difference between the NNPDF and ansatz-1 distributions in the $0.09 \lesssim x \lesssim 0.2$ region.  We note that the small uncertainties of $\Delta g(x)$ from ansatz-1 and ansatz-2 are  biased by the parametrization form, Eqs~\eqref{a1} and~\eqref{a2}, where the $(1-x)$ power difference are fixed between $g^+(x)$ and $g^-(x)$. For the ansatz-3, we introduce independent parameters for the $(1-x)$ powers and the polynomial parts of the two helicity basis distributions. Thus, the ansatz-3 is a more flexible parametrization than ansatz-1 and ansatz-2, but it requires a simultaneous fit to both unpolarized and polarized distributions to determine the parameters. Therefore, the result from ansatz-3 is driven by global analysis and less biased. The difference between the result from ansatz-3 and the one from ansatz-1 or ansatz-2 indicates the model uncertainty of imposing the $(1-x)$ power difference of $g^+(x)$ and $g^-(x)$, or, in other words, how much the uncertainties of the results from ansatz-1 and ansatz-2 are biased.

 Due to the current precision of experimental data, the phenomenological determination of $\Delta G$ is sensitive to the parametrization form in the global analysis. If allowing a possible sign-change of $\Delta g(x)$ at some $x$ value, one will find large uncertainties of $\Delta g(x)$ and thus very poor constraint on $\Delta G$. In our approach,  the helicity retention is incorporated in our parametrization  of ansatz-1 and ansatz-2, where the polarized gluon distribution is  fixed once the unpolarized distribution  is determined.  As we will show in the next section, the result from the ansatz-3 is also consistent with the helicity retention, although it is not imposed in the parametrization form. 

  One can observe that the uncertainties of the $x\Delta g(x)$ determined from ansatz-1 and ansatz-2 are highly constrained. This is due to the bias of the parametrization form, which assumes a relation between the two helicity-basis distributions and thus leads to an underestimation of the uncertainties of the polarized distribution. On the other hand, ansatz-3 is more flexible and the uncertainty of $x\Delta g(x)$ is governed by the global analysis of $x\Delta g(x)$.  An outstanding question is how to distinguish between these  three different  determinations of $x\Delta g(x)$ distributions,  especially in the large $x$-region which is of primary interest for the nonperturbative LQCD calculations of PDFs. One answer is, as we will see in Section~\ref{gluonITD}, the gluon helicity $\Delta G$ obtained from the Ioffe-time distribution  obtained from ansatz-1  parametrization has a magnitude almost twice as large compared to the one resulting from ansatz-2.  Similarly, the Ioffe-time distribution of the polarized gluon distribution obtained from ansatz-1 is almost double in magnitude compared to that obtained from ansatz-2. The difference of these two Ioffe-time distributions in the small Ioffe-time region $\om\approx 0-6$  can be investigated in  LQCD calculations to discriminate between these two ans{\"a}tze. On the other hand, the polarized gluon distribution determined by fitting the NNPDFpol1.1 global analysis~\cite{Nocera:2014gqa} is data-driven and has much larger uncertainty compared to that obtained from ansatz-1 and ansatz-2. Therefore, the resulting polarized ITD also has larger uncertainty. It will be a good opportunity to explore the polarized gluon ITD in LQCD calculations and have a possible impact on constraining the uncertainty. We will present  detailed discussion  of this prospect in Sections~\ref{gluonITD} and~\ref{gluonITD}.


\section{Gluon asymmetry distribution}
The COMPASS experiment at CERN has measured gluon asymmetry distribution $\Delta g(x)/g(x)$ from the cross section helicity asymmetry of photon-gluon fusion $(\gamma^*g\to q\bar{q})$ in the semi-inclusive deep-inelastic scattering (DIS) of proton-proton collision~\cite{Ageev:2005pq,Adolph:2012ca,Adolph:2015cvj}.  Although the open charm events provide the cleanest signal to the $\gamma^*g\to q\bar{q}~(q=c)$ events~\cite{Watson:1981ce,Gluck:1988uj}, the rate of these events is very small. The high statistics two-jets events with large transverse momentum $p_T$   with respect to the virtual photon direction can give access to the photon-gluon fusion subprocess but with a price of significant  background  which  has to be subtracted in a model-dependent way  to determine $\Delta g(x)/g(x)$.

\begin{figure}[htp]
		\begin{center}
			\includegraphics[width=3.45in, height=2.5in]{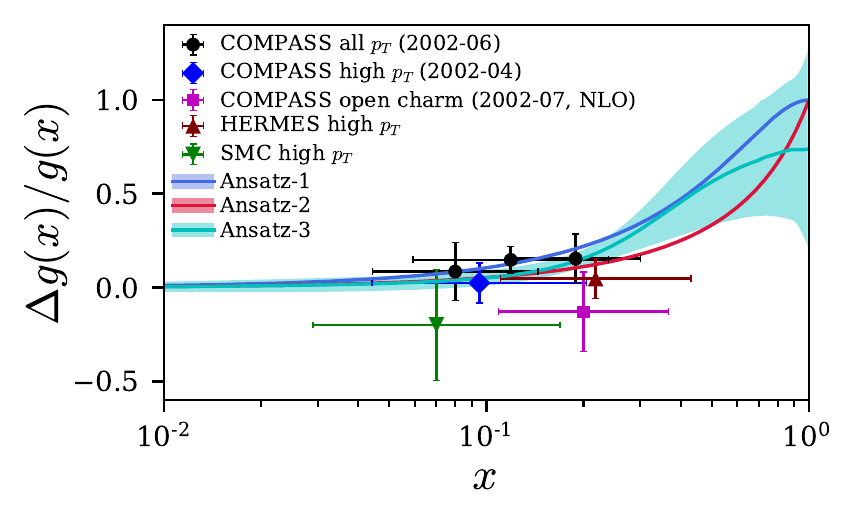}
			\caption{\label{fig:3}
				 Comparison between the two determinations of $\Delta g(x)/g(x)$ from this calculation with the experimental measurements. The direct measurements of COMPASS~\cite{Ageev:2005pq,Adolph:2015cvj}, HERMES~~\cite{Airapetian:2010ac}, and SMC~\cite{Adeva:2004dh} are obtained in leading order from high $p_T$ hadrons and from open charm muon production at COMPASS\cite{Adolph:2012ca} in next-to-leading order at different $x$-values are shown. The blue, red, and cyan bands labeled by Ansatz-1, Ansatz-2, and Ansatz-3 show the gluon asymmetry distributions determined using the parameters obtained using ansatz-1, ansatz-2, and ansatz-3 for the $xg^+(x)$ and $xg^-(x)$ distributions, respectively as discussed in the main text.}
		\end{center}
\end{figure}

In FIG.~\ref{fig:3}, we compare the $\Delta g(x)/g(x)$ ratio obtained from our calculation with data at different $x$-values extracted from high $p_T$ hadrons in the leading-order analyses~\cite{Ageev:2005pq,Adolph:2015cvj} and from the open charm production in the next-to-leading order analysis~\cite{Adolph:2012ca}  at COMPASS, from high $p_T$ hadrons in the leading-order analyses by the Spin Muon Collaboration (SMC) at CERN~\cite{Adeva:2004dh} and at the HERMES experiment~\cite{Airapetian:2010ac}.  We note that the endpoint values of $\Delta g(x)/g(x)$ are fixed in ansatz-1 and ansatz-2. In the limit $x\to 0$ the ratio goes to $0$ and as $x\to1$ the ratio goes to $1$, no matter what values are assigned to the parameters in Eqs.~\eqref{a1} and~\eqref{a2}. The difference between the results from ansatz-1 and ansatz-2 may be regarded as the model uncertainty, while the uncertainty for either of them is model biased.  It is not a surprise to find that the result from ansatz-3 has much larger uncertainty, because the parametrization of ansatz-3 is more flexible and thus less biased. One can also notice that the ratio does not necessarily go to $1$ in the limit $x\to 1$ for the ansatz-3, since we introduce independent parameters for the $(1-x)$ powers of $g^+(x)$ and $g^-$ without the requirement of any $(1-x)$ power suppression of $g^-(x)$ in comparison with $g^+(x)$. However, the result from the simultaneous fit is still consistent with $1$ at the $x\to1$ endpoint. 

The two different solutions of $\Delta g(x)$ obtained in the COMPASS next-to-leading order analyses~\cite{Adolph:2015saz,Wilfert:2017mvt}  allow $\Delta g(x)/g(x)$  to be positive or negative in the entire $x$-region, whereas our analyses with ansatz-1 and ansatz-2 give a positive $\Delta g(x)/g(x)$ in the entire $x$-region and a very small $\Delta g(x)/g(x)$ in the $x<10^{-1}$ region. The $x\to 1$ value of this asymmetry distribution is consistent with the pQCD prediction of the helicity retention~\cite{Farrar:1975yb,Brodsky:1994kg}.  Even for the ansatz-3, where we do not require the helicity retention in the parametrization, the simultaneous fit result driven by the global analysis is still consistent with this prediction.

\section{Gluon Ioffe-time distributions} \label{gluonITD}

As first discussed by Gribov, Ioffe, and Pomeranchuck~\cite{Gribov:1965hf}, the large coherent length distances defined in the target rest frame become important at high energies for the virtual photon-nucleon scattering.  Ioffe further demonstrated a connection between the DIS scattering amplitude and the spacetime representation of the correlator of the electromagnetic current~\cite{Ioffe:1969kf},  establishing  a  relation between longitudinal coherent distances and  Bjorken scaling. Application of the Ioffe-time distribution (ITD) to study parton distributions in  coordinate space using nonperturbative method was proposed in~\cite{Braun:1994jq}. It was argued in~\cite{Braun:1994jq} that calculations of PDFs in  momentum space receive contributions from both small and large longitudinal distances for each value of $x$ and  can result in a problem of treating different physics associated with different distances simultaneously. The Ioffe-time $\tau_I$  measures the interval between the absorption and emission of virtual-photon by a hadron in  DIS and gives the coherence length of the pair production in the target rest frame, 
\bea
\tau_I =\frac{\om}{M}
\eea
where $M$ is the target mass.  Braun {\it et. al.}  entitled the Lorentz invariant variable $\om$ as  the Ioffe-time;  and we will use this naming convention  for the remainder of this article, where the same language is seen in contemporary  coordinate space LQCD formalisms  used to isolate parton distributions~\cite{Braun:2007wv,Radyushkin:2017cyf,Ma:2017pxb}. We  note  $\om= M \tau_{I}$ is  defined in  the hadron's rest frame, hence the designation ``time" despite  ``$\om$" itself  being  neither time nor space in LQCD calculations. It is indeed the $\om$-dependence of  the ITD that converts into the $x$-dependence of the parton distributions. Recently, a method for obtaining frame-independent,  three-dimensional light-front coordinate-space wave functions and its relevance to LQCD calculations of PDFs has also been discussed in terms of frame-independent longitudinal distance (the Ioffe-time) in~\cite{Miller:2019ysh}.

\begin{figure}[htp]
		\begin{center}
			\includegraphics[width=3.35in, height=2.5in]{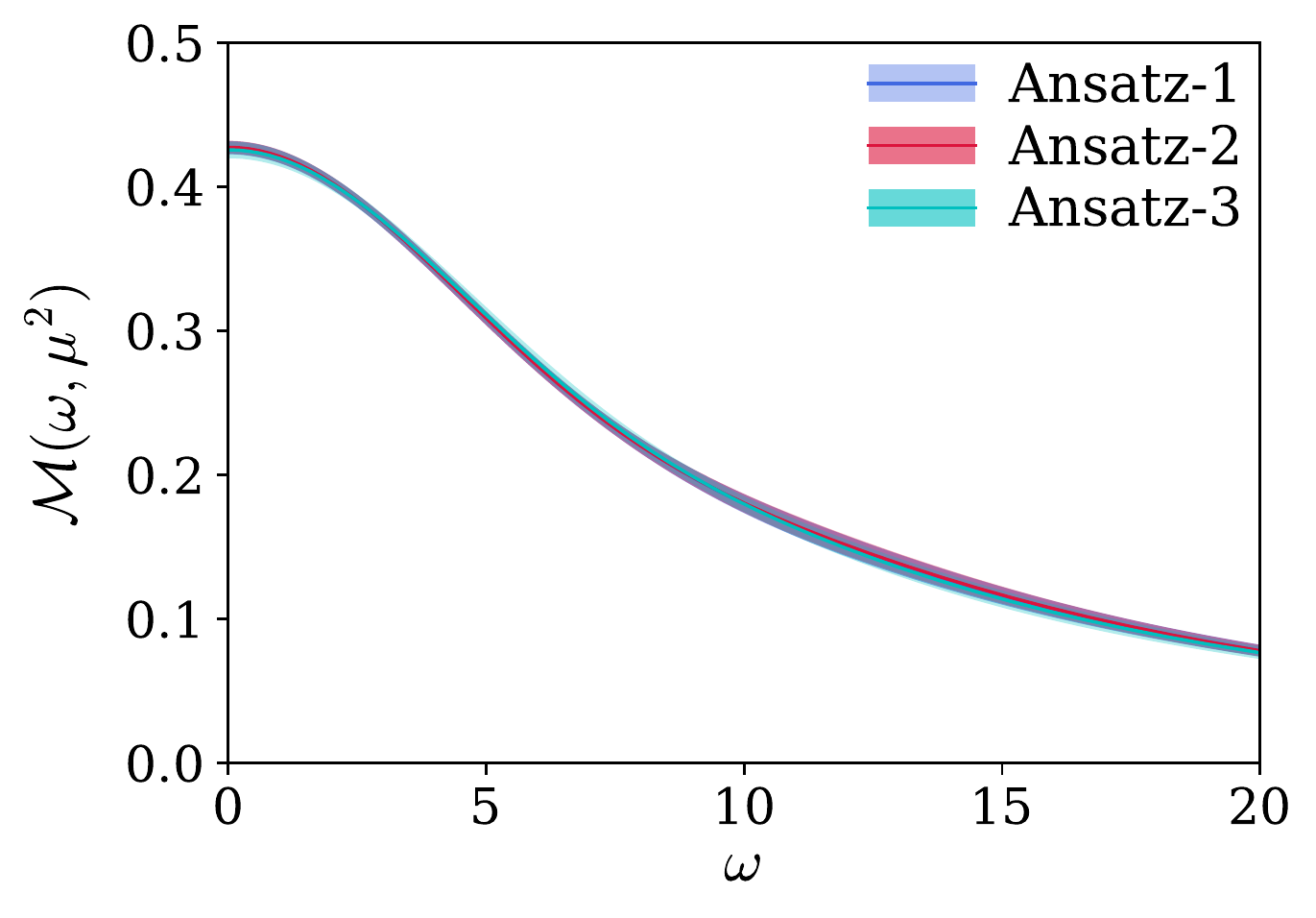}
			\caption{\label{fig:4}
				 Determination of the Ioffe-time distribution of the unpolarized gluon distribution. The blue, red, and  cyan bands, labeled by Ansatz-1, Ansatz-2 and Ansatz-3, show the Ioffe-time distributions determined using the fit parameters according to ansatz-1, ansatz-2, and ansatz-3 for the $xg^+(x)$ and $xg^-(x)$ distributions, respectively. $\mu^2$ indicates the factorization scale associated with the NNPDF unpolarized gluon distribution used in this work.}
		\end{center}
\end{figure}

\begin{figure}[htp]
		\begin{center}
			\includegraphics[width=3.35in, height=2.5in]{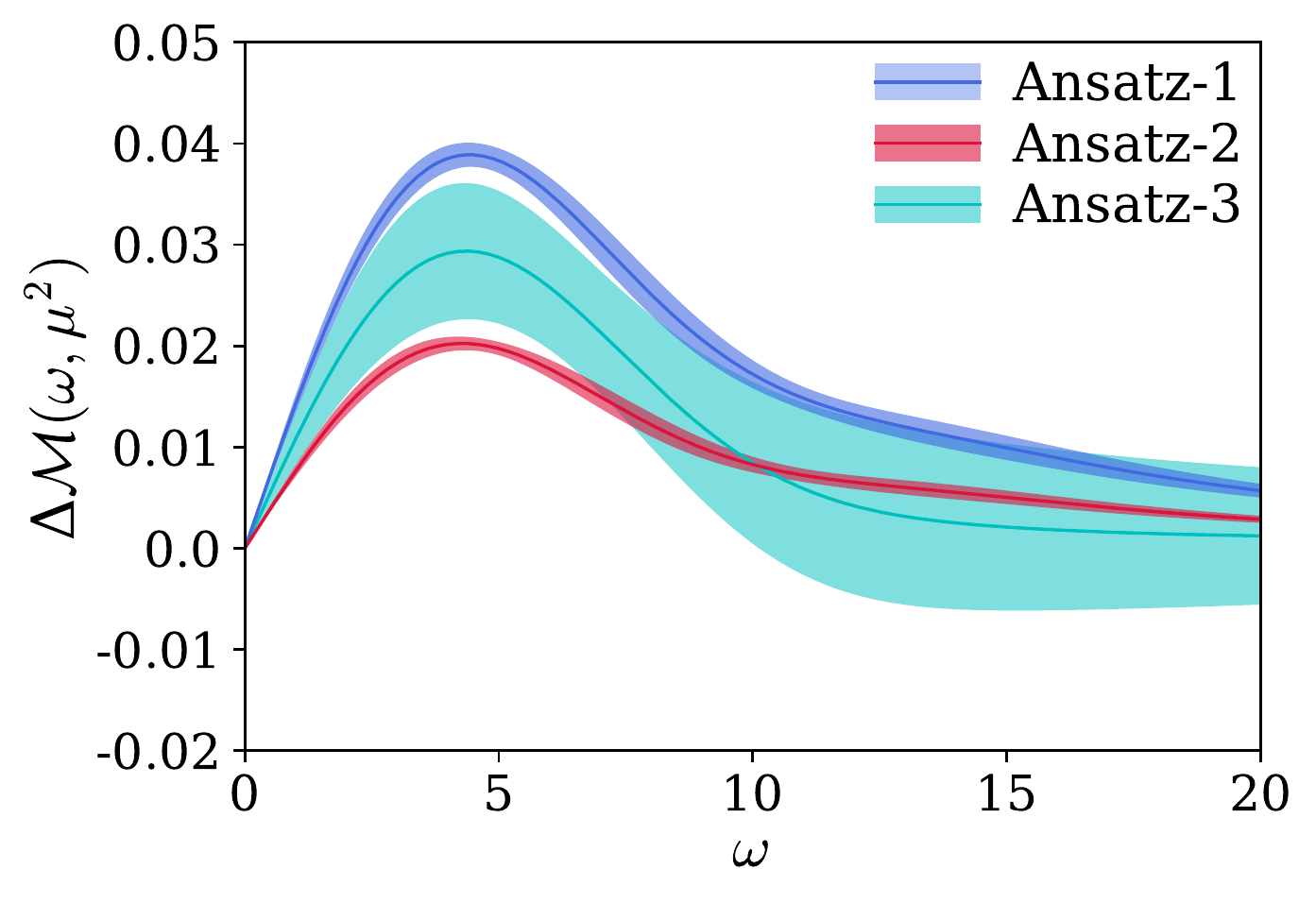}
			\caption{\label{fig:5}
				  Ioffe-time distribution of the polarized gluon distribution labeled by Ansatz-1,  Ansatz-2, and Ansatz-3. The factorization scale $\mu^2$ associated with the NNPDF  gluon distribution used to determine the fit parameters of the $xg^+(x)$ and $xg^-(x)$ distributions is also indicated.}
		\end{center}
\end{figure}

One can now write the unpolarized gluon distribution in terms of its Ioffe-time distribution as~\cite{Braun:1994jq,Saalfeld:1997uv}
\bea \label{Munpol}
\mathcal{M}(\om,\mu^2)&=& \int_0^1 dx\,xg(x,\mu^2)\cos(x\om)
\eea
at a scale $\mu^2$. Similarly using the definition of the polarized gluon distribution from~\cite{Manohar:1991ux}, $x\Delta g(x,\mu^2)$ is related to  its Ioffe-time distribution~\cite{Braun:1994jq,Saalfeld:1997uv}  via
\bea \label{Mpol}
\Delta\mathcal{M}(\om,\mu^2)&=& \int_0^1 dx\,x\Delta g(x,\mu^2)\sin(x\om).
\eea
 In comparison with Eq.~\eqref{Munpol}, the $\sin(x\om)$ in the integrand of Eq.~\eqref{Mpol} leads to one additional power of $x$ suppression when $x$ is small and therefore reduce the small-$x$ region contribution, as well as its uncertainty, to the $\Delta \mathcal{M}(\om,\mu^2)$. With knowledge of the polarized gluon ITD, from Eq.~\eqref{Mpol}, one can immediately obtain the gluon helicity contribution to the nucleon spin 
\bea \label{deltaG}
\Delta G(\mu^2) &=& \int_0^\infty d\om~\Delta\mathcal{M}(\om,\mu^2)\nn \\
&=& \int_0^1 dx~\int_0^\infty d\om~x\Delta g(x,\mu^2) ~{\rm Im}\big(e^{ix\om}\big)\nn \\
&=&\int_0^1 dx~\Delta g(x,\mu^2),
\eea
 where we have used the principal value prescription to calculate the integral $\int_0^\infty d\om~\sin(x\om)$. As seen from Eq.~\eqref{deltaG}, and  as will be explored further in Section~\ref{asymp}, we shall see that access to the asymptotic region of the ITD $\Delta\mathcal{M}(\om,\mu^2)$, namely up to $\om\approx 15$ can provide a stringent constraint on the gluon helicity in the nucleon. 
Given the gauge-invariant and frame-independent definition of the ITD, one can take advantages of calculating the $\Delta G(\mu^2)$ from the ITD to avoid the issues in the spin-decomposition~\cite{Leader:2013jra}. 
Using Eqs.~\eqref{Munpol} and~\eqref{Mpol}, we calculate the ITDs of the unpolarized and polarized gluon distributions from the parameterizations of $g^{+}(x)$ and $g^{-}(x)$ and present them in FIGs.~\ref{fig:4} and~\ref{fig:5}, respectively.

 Similar to the unpolarized gluon PDF, we see from FIG.~\ref{fig:4} that ans{\"a}tze  1, 2, and 3 produce almost identical ITDs for the $xg(x)$ distributions. However, we see that the magnitude of the $x\Delta g(x)$ ITD (FIG.~\ref{fig:5}) in the $\om\sim 5-10$ range is almost double when using the ansatz-1  parametrization relative to the ansatz-2  parametrization.  Consequently, using Eq.~\eqref{deltaG}, we obtain $\Delta G= 0.451(7)$ from ansatz-1 and $\Delta G= 0.258(4)$ from ansatz-2.  As mentioned above, the difference between the $\Delta G$ values from ansatz-1 and ansatz-2 may be viewed as model uncertainty, while the uncertainty band of either one is biased by the parametrization form. The result from the simultaneous fit with ansatz-3 is $\Delta G= 0.23(41)$. Such a large uncertainty from ansatz-3 indicates the fact that the $\Delta G$ is still very poorly known. We note that the lower value of $\Delta G= 0.258(4)$ obtained from ansatz-2 is  consistent with the previous LQCD determination~\cite{Yang:2016plb} of $\Delta G$. The most recent calculation of the nucleon spin decomposition at the physical pion mass~\cite{Alexandrou:2020sml} found  the total gluon angular momentum contribution in the proton to be $0.187(47)$. According to this calculation, unless the gluons contribute a large and negative orbital angular momentum to the nucleon total angular momentum budget, the $\Delta G$ contribution to the proton spin is expected to be less than $\sim 50\%$. On the other hand, if one excludes the low $x$ contribution of the polarized gluon PDF one obtains $\Delta G\sim 40\%$  for  $x>0.02$ in~\cite{Nocera:2014gqa} and  for $x>0.05$ in~\cite{deFlorian:2014yva}. Of course, these values can change in global fits, if gluons  are shown to have large positive contributions to the nucleon spin - a prospect to be explored at the EIC~\cite{Accardi:2012qut}. Another possibility is the $x\Delta g(x)$ distribution can change sign in the low-$x$ region,   thereby reducing the contribution of $\Delta G$ to the nucleon spin budget.

One important question we encountered in Sec.~\ref{gluonPDFs} was  how one can discriminate between ansatz-1 and ansatz-2 and the resulting $\Delta g(x)$ distributions.  An  interesting  feature  we observe from FIGs.~\ref{fig:4} and~\ref{fig:5} is  the significant difference in the magnitude between these two ITDs in the $\om\sim 0-5$ range.  This will provide a great opportunity for  LQCD calculations to discriminate between these two different  determinations for the $x \Delta g(x)$ distribution.  On the other hand, the polarized gluon ITD obtained from ansatz-3 has a much larger uncertainty. It remains to see that if LQCD calculation in the lower Ioffe-time region can be precise enough to provide complementary information about the uncertainty of the polarized gluon ITD. We highlight  recent LQCD calculations~\cite{Sufian:2019bol,Sufian:2020vzb,Joo:2020spy,Bhat:2020ktg,Fan:2020cpa}  have obtained  precise  ITD data in the $\om\lesssim 5$ region.  Resolution to  this problem in  future LQCD calculations of the polarized gluon ITD  hinges on precise LQCD data for $\om\sim 0-5$, as well as mitigation of higher-twist effects which in turn ensures the validity of the short-distance pQCD factorization of lattice matrix elements into PDFs.

\section{Calculation of higher moments from gluon Ioffe-time distributions}\label{moments}
\begin{figure}[htp]
		\begin{center}
			\includegraphics[width=3.45in, height=2.5in]{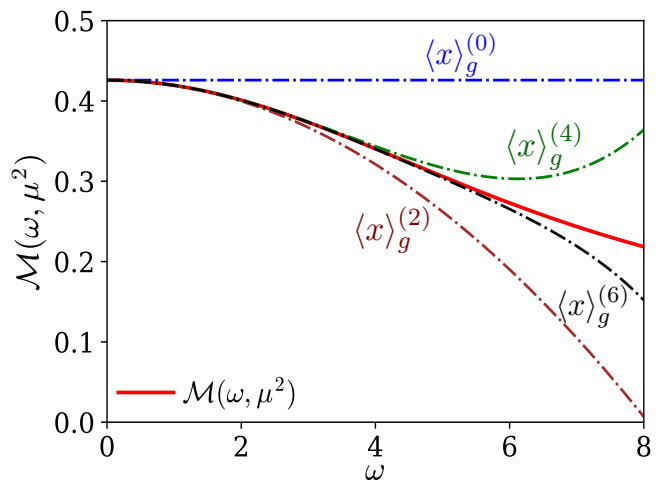}
			\caption{\label{fig:6}
				 Approximation of the unpolarized gluon Ioffe-time distribution $\mathcal{M}(\om,\mu^2)$ using  moments of the distribution. The cyan band denotes the Ioffe-time distribution determined using the fit parameters of $xg^+(x)$ and $xg^-(x)$ for ansatz-2 from 1 replica  of  the  unpolarized gluon  distribution  from  NNPDF3.1 NLO PDF. The lines labeled by $\langle x\rangle_g^{(n)}$ denote an approximation which require the knowledge of the first $n$ nonvanishing moments of the unpolarized gluon distribution to reproduce the ITD in increasingly larger range of $\om$.}
		\end{center}
\end{figure}

From the ITDs in Eqs.~\eqref{Munpol} and~\eqref{Mpol}, one can immediately calculate the associated higher moments of the distributions. Performing a power series expansion of $\cos(x\om)$ in Eq.~\eqref{Munpol} (with scale $\mu^2$ omitted ),

\bea \label{unpolmoments}
\mathcal{M}(\om) &=& \int_0^1dx~xg(x)\sum_{n=0}^{\infty}(-1)^n\frac{(x\om)^{2n}}{(2n)!}\nn \\
&=& \langle x\rangle_g^{(0)} -\frac{\om^2}{2!} \langle x\rangle_g^{(2)} + \frac{\om^4}{4!} \langle x\rangle_g^{(4)} - \cdots 
\eea
we can get access to any number of moments depending on the available   $\om$-range of ITDs.  It is clearly seen from FIG.~\ref{fig:6} that  one can reproduce the ITDs with increasingly better accuracy and in the larger $\om$-region in terms of higher moments. Another way to state this observation is that, access to ITD in increasingly larger $\om$ range would result in access to increasingly higher number of moments. However, even without access to ITD in the region $\om\to \infty$, it is still possible to obtain PDFs  with very good accuracy (again assuming the Regge phenomenology is valid in approximately $x>10^{-2}$ region). As we will see in the next Section that access to ITD in the region $\om\approx 0-15$ for the gluon distributions can be sufficient to extract their $x$-dependence in the $10^{-2}>x>1$ region.   

Similarly, from Eq.~\eqref{Mpol}, one can express the polarized gluon ITD in terms of odd moments as
\bea\label{polmoments}
\Delta\mathcal{M}(\om) =  \om\langle\Delta x\rangle_g^{(3)}  -\frac{\om^3}{3!} \langle\Delta x\rangle_g^{(5)}  + \frac{\om^5}{5!} \langle\Delta x\rangle_g^{(7)} -  \cdots\nn \\
\eea
From Eqs.~\eqref{deltaG} and~\eqref{polmoments} it is seen that a calculation of ITD of polarized gluon distributions in $\om\sim 15$ region will not only give a reliable estimate for gluon helicity $\Delta G$ in the nucleon but also several other higher moments which are currently unknown in theoretical calculations and not constrained by the experimental data.

\begin{figure}[htp]
		\begin{center}
			\includegraphics[width=3.45in, height=2.35in]{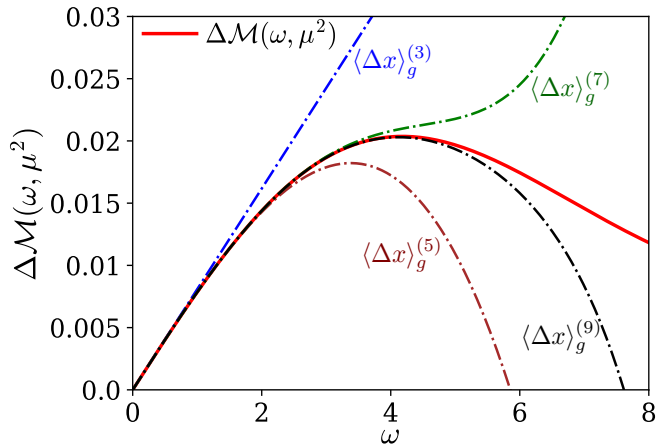}
			\caption{\label{fig:7}
				  The polarized gluon Ioffe-time distribution $\Delta\mathcal{M}(\om,\mu^2)$ approximated by its $n$-th order Taylor expansion around $\om=0$, determined from the fit paraemters for ansatz-2 from 1 replica  of  the  unpolarized gluon  distribution  from  NNPDF3.1 NLO PDF. The lines labeled by $\langle \Delta x\rangle_g^{(n)}$ denote an approximation which require the knowledge of the first $n$ nonvanishing moments of the polarized gluon distribution to reproduce the ITD in increasingly larger range of $\om$.}
		\end{center}
\end{figure}

We see from FIGs.~\ref{fig:6} and~\ref{fig:7} that with increasing number of moments according to the Taylor expansion in Eqs.~\eqref{unpolmoments} and~\eqref{polmoments}, the unpolarized and polarized gluon distribution can be approximated in the increasing range of Ioffe-time $\om$.  The accurate reproduction of the ITDs in terms of higher and higher moments can be understood from the fact that  the Taylor expansion of $\mathcal{M}(\om)$ about $\om=0$ has infinite radius of convergence and therefore even the asymptotic region can be continuously reached from the origin of $\mathcal{M}(\om)$ as was also demonstrated in~\cite{Mankiewicz:1996ep}. Therefore along with the advantage that one can perform a reliable extraction of PDFs from the ITDs given the precision, accuracy and availability of ITD over a moderate range of $\om\lesssim 15$, this also serves as a powerful formalism to calculate higher moments which are not feasible in the LQCD calculations due to power divergent mixing of lower mass dimension operators in the local moments calculations.

 FIGs.~\ref{fig:6} and~\ref{fig:7} illustrate that access to large-$\om$ region is necessary to obtain precise higher moments from LQCD calculations.  Although the asymptotic limits of ITDs are dominated by the small-$x$ physics, access to large-$\om$ region is also important for a precise determination of large-$x$ physics, as we will see from the derivation of the asymptotic limits of gluon ITDs in the next section.

\section{Asymptotic limit of gluon Ioffe-time distribution} \label{asymp}
\begin{figure}[htp]
		\begin{center}
			\includegraphics[width=3.5in, height=2.4in]{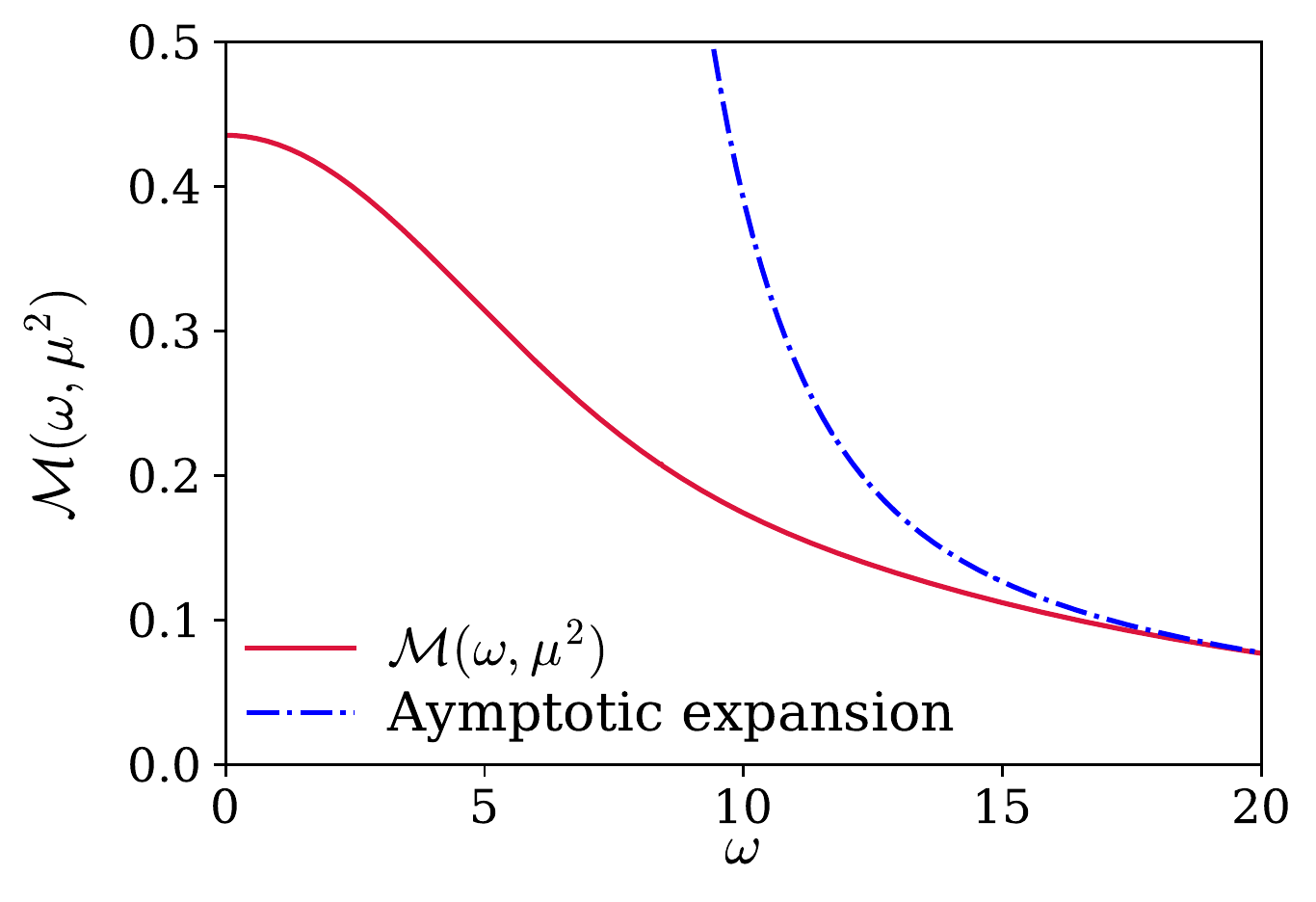}
			\caption{\label{fig:asunpol}
				 Asymptotic expansion of the unpolarized gluon ITD corresponding to a given set of parameters  obtained by fitting one replica of NNPDF unpolarized gluon distribution using ansatz-2 for the $xg^+(x)$ and $xg^-(x)$ distributions. The cyan line indicates the ITD and the dashed line indicates the asymptotic limit of the ITD governed by the corresponding fit parameters. }
		\end{center}
\end{figure}

\begin{figure}[htp]
		\begin{center}
			\includegraphics[width=3.5in, height=2.4in]{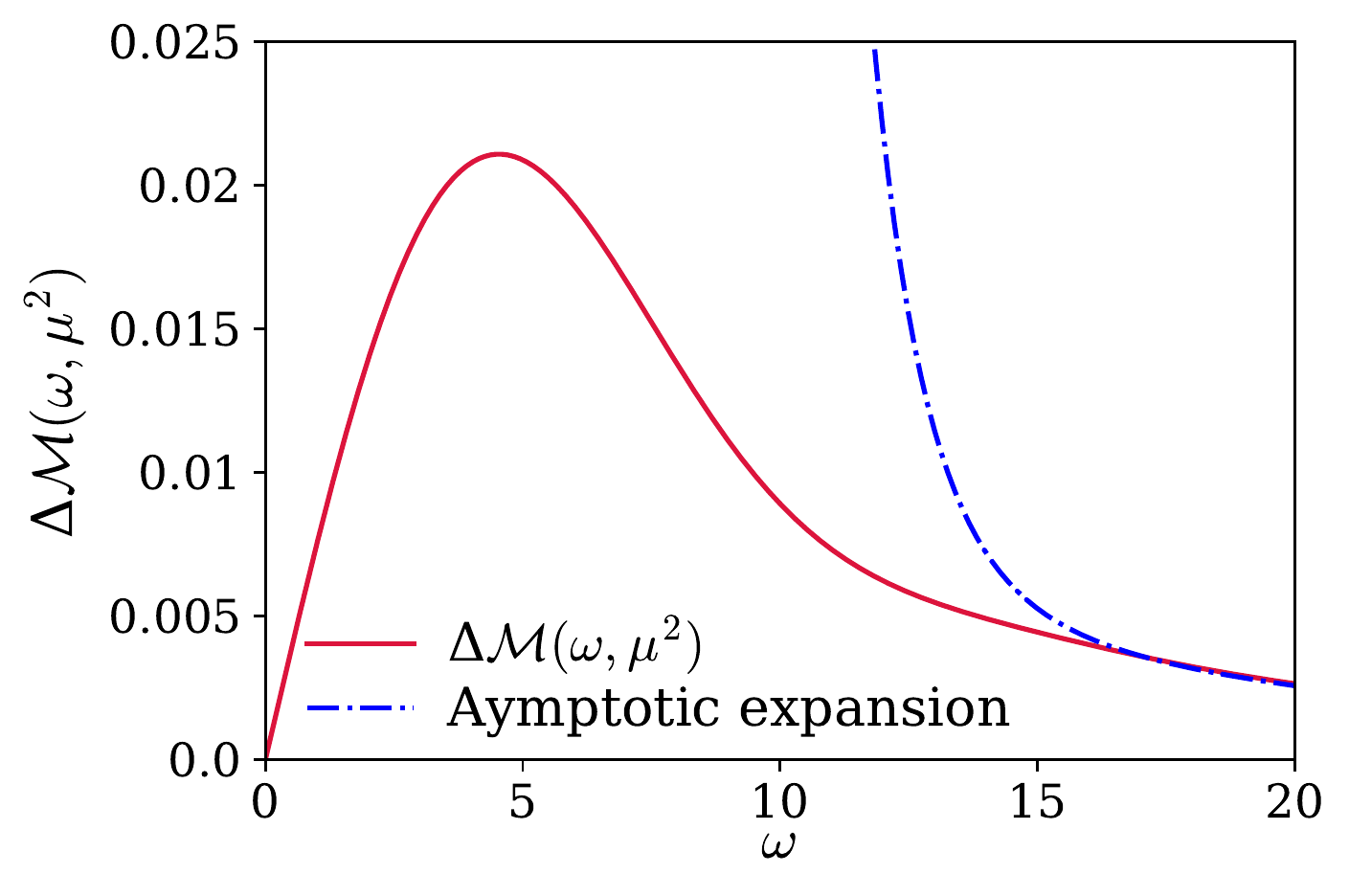}
			\caption{\label{fig:aspol}
				 Asymptotic expansion of the polarized gluon ITD corresponding obtained by fitting one replica of NNPDF unpolarized gluon distribution using ansatz-2. The dashed line indicates the asymptotic limit of the polarized gluon ITD corresponding to fit parameters.}
		\end{center}
\end{figure}

In this section, we calculate  analytic expressions for the asymptotic limits of the unpolarized and polarized gluon ITDs associated with the functional forms of $xg(x)$ and $x\Delta g(x)$ used in Section~\ref{gluonPDFs}.  We examine the $\om$ region around which one can approach the asymptotic region of the ITDs.  We again emphasize that the purpose of this calculation is not to calculate the gluon PDFs at extremely small $x$ and we therefore consider only the low-$x$ region where the Regge phenomenology is valid and avoid the complication with unresolved issues in the extremely small $x<10^{-3}$ region~\cite{McLerran:1993ni,JalilianMarian:1996xn,Kuraev:1977fs,Balitsky:1978ic,Kirschner:1983di,Gribov:1984tu,Mueller:1985wy,Balitsky:1995ub,Kovchegov:1999yj}. To calculate the asymptotic limits of $\mathcal{M}(\om)$ and $\Delta \mathcal{M}(\om)$, we start with the simplest functional form of PDFs $x^a(1-x)^b$ since $xg(x)$ and $x\Delta g(x)$ determined using functional forms in Eqs.~\eqref{a1} and~\eqref{a2} can  be viewed as linear combination of this form with different values of exponents and appropriate normalizations. We first  consider the asymptotic expansions (in the limit $\omega \rightarrow \infty$) of the following integral:
\bea
\int_0^1  x^a (1-x)^b  \exp{( i \omega x)} \, dx &=& E_{R} ( a,b;\,\omega) \nn \\
&+& \mathcal{O} \left(1 / \omega^{a+R+1}\right),
\eea
where $E_{R} (a,b;\,\omega)$ is the standard aysmptotic expansion of the confluent hypergeometric function, $ M(a+1, a+b+2; \, i\omega)$, and up to order $\omega^{-a-R}$:
\bea
  E_{R} ( a,b;\,\omega) &\equiv & \sum_{n = 0} ^{R-1} \frac{\Gamma(a+n+1) (-b)_n}{n!} \bigg(\frac{i}{\omega}\bigg)^{a+n+1} \nn \\ 
  & +& \sum_{n = 0} ^{R + a -b -1} \frac{\Gamma(b+n+1) (-a)_n}{n!} \bigg(\frac{i}{\omega}\bigg)^{b+n+1}.\nn \\
\eea
We also have
\bea
\int_0^1  x^a (1-x)^b  \cos{( \omega x)} \, dx &=& \operatorname{Re} \big( E_{R} ( a,b;\,\omega)\big) \nn \\ 
& +& \mathcal{O} \big(1 / \omega^{a+R+1}\big),
\eea
and
\bea
\int_0^1  x^a (1-x)^b  \sin{( \omega x)} \, dx &=& \operatorname{Im} \big( E_{R} ( a,b;\, \omega)\big) \nn \\ 
& +& \mathcal{O} \big(1/\omega^{a+R+1}\big) 
\eea
\newpage
\noindent We define,
\begin{align}
  C_{R} ( a,b;\,\omega) &\equiv \operatorname{Re} \left( E_{R} ( a,b;\,\omega)\right), \\
    S_{R} ( a,b;\,\omega) &\equiv \operatorname{Im} \left( E_{R} ( a,b;\,\omega)\right)
\end{align}
Using the above expressions, we can summarize the asymptotic expansions  of the unpolarized and polarized gluon ITDs for ansatz-1:
\bea
&&\mathcal{M}(\om,\mu^2) \nonumber =   A ~\big[ ~\big(~C_R (\alpha , 4+\beta;\,\omega) \nn \\
&&+ \gamma \, C_R (\alpha + 1/2 , 4+\beta;\,\omega) + \delta\, C_R (\alpha +1 , 4+\beta;\,\omega)~ \big)\nn \\
&&+ \big(~\beta \rightarrow \beta + 2 ~\big) ~\big]  +  B ~\big[ ~\beta \to \beta + 1 ~\big] \nn \\
&& + \mathcal{O} \big(1 / \omega^{a+R+1}\big)
\eea 
and
\bea
&&\mathcal{M}(\om,\mu^2) =   A ~\big[ ~\big(~S_R (\alpha , 4+\beta;\,\omega) \nn \\
&&+ \gamma \, S_R (\alpha + 1/2 , 4+\beta;\,\omega) + \delta\, S_R (\alpha +1 , 4+\beta;\,\omega)~ \big)\nn \\
 &&- \big(~\beta \to \beta + 2 ~\big) ~\Big] +  B ~\big[ ~\beta \to \beta + 1 ~\big] \nn \\
 &&+ \mathcal{O} \big(1 / \omega^{a+R+1}\big)
\eea
 Similar expressions can be written for ansatz-2 and ansatz-3. We show the asymptotic limits of the unpolarized and polarized gluon ITDs for  ansatz-2 in FIGs.~\ref{fig:asunpol} and~\ref{fig:aspol}, respectively.  For the demonstration purpose, we select one arbitrary set of parameters from to the fit to 1 replica  of  the  NNPDF unpolarized gluon  distribution described in Section~\ref{gluonPDFs}.

The $\mathcal{M}(\om)$ and $\Delta\mathcal{M}(\om)$ approach the asymptotic limits around $\om\sim 15$ as can be seen in FIGs.~\ref{fig:asunpol} and~\ref{fig:aspol}. It is important to note that  if future LQCD calculations of gluon ITD can reach the region $\om\sim 15$, they will be able to provide nonperturbative information to the Lipatov's pomeron~\cite{Kuraev:1977fs,Balitsky:1978ic}.

Using the fact that gluon PDF diverges much faster than the valence quark PDF in the limit $x\to 0$, one can  show that the asymptotic limit of the ITD corresponding to nucleon valence quark distribution will set in at  earlier $\om$ compared to the gluon ITDs, also noted in~\cite{Braun:1994jq}. This implies that the asymptotic region of the nucleon valence quark ITD can be approached easily in the nonperturbative calculations compared to the gluon ITDs.

\section{Applications to lattice QCD calculations of PDFs}\label{LQCD}
In recent years, several LQCD methods have been proposed and  developed to probe the light-cone structure of hadrons, including  the path-integral  formulation  of  the  deep-inelastic  scattering  hadronic  tensor~\cite{Liu:1993cv},  coordinate-space method for the calculating light-cone distribution amplitudes~\cite{Braun:2007wv},  inversion method~\cite{Horsley:2012pz}, quasi-PDFs/LaMET~\cite{Ji:2013dva,Ji:2014gla}, pseudo-PDFs~\cite{Radyushkin:2017cyf}, and good lattice cross sections~\cite{Ma:2014jla,Ma:2017pxb}.  For the most recent review of LQCD calculations of PDFs, see~\cite{Lin:2020rut}. 

The extraction of PDFs from LQCD calculations has received great interest since Ji's proposal in~\cite{Ji:2013dva,Ji:2014gla}. Instead of directly calculating the light-cone correlation functions which define the PDFs, one can extract them from the spatial correlation  of parton fields  calculable on  the Euclidean lattice. We begin this section by acknowledging that any LQCD calculations of PDFs using any  of the above formalisms share the common challenge of how best to extract a continuous distribution from discrete data, compounded by a limited  number of data points due to a finite range of spatial separations and hadron momenta. As the ITDs in question herein are Fourier transforms of the underlying PDFs, the available data in a discrete and limited domain  leads to an ill-posed inverse problem well-known  to the LQCD community~\cite{Karpie:2019eiq,Liang:2019frk,Bhat:2020ktg,Alexandrou:2020tqq}. Similar to functional forms  used in global fits, different PDF parametrizations obviate  the ill-posed inverse Fourier transform  at the cost of  additional systematic errors in the determination of PDFs. The analyses of the global fitting community have matured over the past several decades to where these systematics can be reliably estimated.  In the ideal scenario,  precise LQCD data  will allow  this systematic error  to likewise  be estimated and corrected by fitting several models and examining  relevant figures of merit.  Significant investigations are being performed  at present to handle this problem better; even the neural network approach for determining PDFs from LQCD data has become feasible now, as we will discuss below.

With the current resources available in LQCD calculations of PDFs, it remains a challenge to obtain precise and accurate data at large $\om=z\cdot p$, where $p$ is the hadron momentum and  $z$ is the spatial separation between parton  fields~\cite{Ji:2013dva,Radyushkin:2017cyf}  or  gauge-invariant currents~\cite{Braun:2007wv,Ma:2017pxb}. This problem was actually anticipated 25 years ago in~\cite{Braun:1994jq}. As a remedy to this problem, it was proposed to consider the low $\om$ region, where LQCD data can be precise and accurate, and the large $\om$ domain of the ITD separately. The LQCD accessible small $\om$-region and the large $\om$ region could then be matched with the knowledge of Regge phenomenology as described in~\cite{Braun:1994jq,Mankiewicz:1996ep,Saalfeld:1997uv}. This approach therefore enables one to obtain reliable estimates of PDFs in the entire $x$-region. Recently, a similar proposal was suggested in~\cite{Ji:2020brr} to circumvent such a problem in the lattice QCD calculation of quasi-PDFs using the LaMET formalism.

First-principles LQCD determinations of the gluon unpolarized and polarized distributions, with controlled statistical and systematic uncertainties, have been of particular interest recently with significant theoretical developments~\cite{Wang:2017qyg,Wang:2017eel,Zhang:2018diq,Li:2018tpe,Balitsky:2019krf}  soliciting a synergy of increasing importance between experimental and theoretical efforts. However, convincing LQCD calculation of gluon PDFs has remained very challenging. It is a big challenge to obtain precise ITD at large $\om$~\cite{Fan:2020cpa} and to reach the asymptotic region of the ITD,  especially for the gluonic observables. In this light, the phenomenological knowledge of the asymptotic limits of the gluon ITDs can be considered as an interesting opportunity to match the low-$\om$ ITD from LQCD calculations and large-$\om$ asymptotic limits utilizing the method discussed in Section~\ref{asymp}. 

LQCD data with larger separations between the quark/gluon fields are seen to be consistent with zero, even change sign, getting affected by various systematic errors due to issues in the renormalization and perturbative matching, and are also expected to be contaminated by higher-twist effects~\cite{Ji:2020brr}. This can be directly seen from the nucleon ITD calculated in the pseudo-PDF formalism and also in the renormalized matrix elements of the quasi-PDF/LaMET formalism both of which start with the same matrix elements~\cite{Ji:2013dva}. This leads to the potential issue with the LQCD calculations of PDFs; the ITD falls off much faster  compared to the ITD constructed from the phenomenologically determined PDFs, e.g. nucleon valence quark distribution which is very well constrained from different global fits. From the phenomenological point of view, this faster-fall of the LQCD data immediately results in a faster-converging PDF at low $x$ and softer fall-off at large $x$. This has been demonstrated using neural network framework  applied to  the extraction of the nucleon valence quark distribution  (which is assumed to be a simpler LQCD calculation  compared to the gluon PDF, sea-quark distributions, etc.) from LQCD calculated matrix elements using the quasi-PDF~\cite{Cichy:2019ebf} and pseudo-PDF~\cite{DelDebbio:2020rgv} approaches.  A similar observation can be found in a recent Monte Carlo based analysis of LQCD data for nucleon valence quark PDF~\cite{Bringewatt:2020ixn}. To investigate these issues, for example, one can construct ITDs from the well-determined nucleon valence PDFs of global fits and compare them with the LQCD calculated matrix elements and examine at which $\om$ the LQCD determined ITD starts to deviate from the phenomenological ITD.     With the help of the asymptotic limit of the ITD, one can also investigate the possible sources of discrepancy between the LQCD calculation and the ITD derived from global fits, such as higher-twist contributions. Investigation based on these observations is an ongoing subject of a future research project.


Along with the calculation of PDFs as discussed above, an ITD can also be used to determine moments of PDFs in a  reliable manner. One can immediately fit the LQCD calculated ITD using Eqs.~\eqref{unpolmoments} and~\eqref{polmoments} to extract moments of PDFs. The accuracy and the number of accessible moments will depend on the available $\om$-range  and on the precision of the ITD. This can provide an alternative approach for extracting local moments of the distribution~\cite{Karpie:2018zaz,Gao:2020ito} by directly  fitting the ITD. In particular, this can be useful for the gluon~\cite{Fan:2020cpa} and sea-quark~\cite{Zhang:2020dkn} distributions   for which the LQCD data is seen to be much noisier compared to the non-singlet quark distributions  and the ill-posed inverse problem of extracting PDFs is much severe. 

\section{Conclusion and Outlook}

 In this  paper, we  investigate the unpolarized and polarized gluon distributions and their applications to the Ioffe-time distributions, which are closely related to the extraction of PDFs from lattice QCD calculations. We parametrize the gluon distributions in the helicity basis and construct the functional form motivated by the counting rules based on perturbative QCD analyses at large $x$ and the phenomenological behavior at low $x$.
Once the helicity-basis gluon distributions are determined, one can easily obtain the unpolarized and polarized gluon distributions. By fixing the $(1-x)$ difference and apply the same polynomial factor, we determine the helicity-basis distributions with the unpolarized gluon distribution and infer the polarized gluon distribution using two different ans{\"a}tze. Although the results are both in relatively good agreement with the global analysis, the two results show a sizable difference from each other. To see the model uncertainty, we also perform a simultaneous fit to unpolarized and polarized gluon distributions from global analyses using a more flexible parametrization form.

As an application, we calculate the Ioffe-time distributions and discuss the possibility that the asymptotic expansion of the Ioffe-time distributions in the large-$\om$ region which can be combined with the future lattice QCD calculations that are limited in a relatively smaller $\om$-range and can guide to extrapolate the lattice data in the larger $\om$-region. Using our calculation, we have demonstrated that the  magnitude of the polarized gluon Ioffe-time distribution within a moderate Ioffe-time window in a future nonperturbative QCD calculation can provide important constraints on the gluon spin content in the nucleon.  We have discussed, as a general application of this method, the discrepancy between the fall-off rate of the phenomenologically well-determined Ioffe-time distributions and those calculated in lattice QCD calculations can serve as an interesting platform to investigate higher twist effects in lattice QCD calculations. Following recent observations, we have discussed that with the present and possibly near future resources and numerical techniques,  lattice QCD calculations alone might not be sufficient enough to extract the full $x$-dependence of PDFs with desired precision and accuracy. In such cases, phenomenologically well-determined Ioffe-time distributions in the large $\om$-region can provide complementary information to the ongoing efforts of the calculation of $x$-dependent hadron structures within first-principles nonperturbative QCD calculations. 

\acknowledgments{RSS thanks Colin Egerer, Joseph Karpie, Nikhil Karthik, Kostas Orginos, and David Richards for useful suggestions and discussions. We thank Jian-Wei Qiu who  provided  insights  and  expertise that greatly assisted this research. RSS is supported in part by U.S. DOE grant No. DE-FG02-04ER41302. TL is supported in part by National Natural Science Foundation of China under Contract No. 11775118. This work is also supported by the U.S. Department of Energy contract  No. DE-AC05-06OR23177,  under  which  Jefferson  Science  Associates, LLC, manages and operates Jefferson Lab,  and  within the framework of the TMD Topical Collaboration. }


\end{document}